\documentclass[12pt]{JHEP3}

\usepackage{amssymb,amsmath}
\usepackage{epsfig}
\usepackage{citesort}

\newcommand{\beq}{\begin{equation}}
\newcommand{\eeq}{\end{equation}}
\newcommand{\baq}{\begin{eqnarray}}
\newcommand{\eaq}{\end{eqnarray}}

\title{The effect of inhomogeneous expansion on the supernova observations}

\author{Kari Enqvist$^{1,2,}$\footnote{E-mail: kari.enqvist@helsinki.fi} ~and Teppo Mattsson$^{1,2,}$\footnote{E-mail: teppo.mattsson@helsinki.fi}\\
${}^1$ Helsinki Institute of Physics, P.O. Box 64, FIN-00014 University of Helsinki, Finland\\
${}^2$ Department of Physical Sciences, P.O. Box 64, FIN-00014
University of Helsinki, Finland}

\abstract{ We consider an inhomogeneous but spherically symmetric
Lemaitre-Tolman-Bondi model to demonstrate that spatial variations
of the expansion rate can have a significant effect on the
cosmological supernova observations. A model with no dark energy but
a local Hubble parameter about $15$ \% larger than its global value
fits the supernova data better than the homogeneous model with the
cosmological constant. The goodness of the fit is not sensitive to
inhomogeneities in the present-day matter density, and our best fit
model has $\Omega_M(r) \sim 0.3$, in agreement with galaxy surveys.
We also compute the averaged expansion rate, defined by the Buchert
equations, of the best fit model and show explicitly that there is
no average acceleration.}

\preprint{HIP-2006-39/TH}


\begin{document}

\section{Introduction}\label{intro}

The simplest homogeneous and isotropic cosmological model within the
framework of general relativity is parameterized by two numbers: the
Hubble constant $H_0$ and the density parameter $\Omega_M$,
interpreted respectively as the average expansion rate and the
average density of non-relativistic matter in the present universe.
This model was generally considered a good description of the
universe on the largest scales until the high redshift supernova
observations in the late 90's \cite{Riess:1998cb}. With the latest
data from supernovae \cite{Riess:2004nr,Astier:2005qq}, galaxy
distributions \cite{Eisenstein:2005su} and anisotropies of the
cosmic microwave background \cite{Spergel:2006hy} the simplest
homogenous and isotropic model would now lead to a highly
contradictory picture of the universe, as the best fit values of
this model for the average matter density demonstrate:

\begin{itemize}
\item Cosmic microwave background: $\Omega_M \sim 1$
\item Galaxy surveys: $\Omega_M \sim 0.3$
\item Type Ia supernovae: $\Omega_M \sim 0 $
\end{itemize}
The natural conclusion is that at least one of the assumptions of
the model must be false.

As is well known, the problem has conventionally been remedied by
introducing the cosmological constant $\Lambda$ or vacuum energy
$\Omega_\Lambda$, giving rise to an accelerated expansion of the
universe. Indeed, the analysis of the cosmological data with both
the vacuum energy and matter components included yields a consistent
picture of the universe, known as the concordance $\Lambda$CDM
model, with the following best fit values for the density
parameters:
\begin{itemize}
\item Cosmic microwave background: $\Omega_M +\Omega_\Lambda \sim 1$
\item Galaxy surveys: $\Omega_M \sim 0.3$
\item Type Ia supernovae\footnote{In the supernova data analysis spatial flatness ($
\Leftrightarrow  \Omega_M +\Omega_\Lambda = 1$) has been assumed.
However, these values are within the $1\sigma$ error of the best fit
values without the flatness constraint \cite{Riess:2004nr}:
$\Omega_M \sim 0.5 $ and $\Omega_\Lambda \sim 1$. }: $\Omega_M \sim
0.3$ and $\Omega_\Lambda \sim 0.7$
\end{itemize}
Moreover, the data seems to require $\Omega_\Lambda > 0$ at a high
confidence level \cite{Riess:2004nr,Astier:2005qq}.

Although the cosmological concordance $\Lambda$CDM model fits all
the observations well, it is plagued by theoretical problems
\cite{Copeland:2006wr}. We do not have a theory that would explain,
not to mention predict, the required value of the cosmological
constant. Various dark energy models, which have been studied
intensively in order to provide a dynamical explanation for the
cosmological constant, are not compelling from the particle physics
point of view and often require fine-tuning
\cite{Copeland:2006wr,Straumann:2006tv}. Modifications of the
general theory of relativity on cosmological scales appear to suffer
from analogous problems; in fact, it has recently been argued that
the cosmological constant seems to be essentially the only
modification that fits all the cosmological data
\cite{Koivisto:2006ie}.

As a rejection of either matter domination or Einstein gravity leads
to trouble, it is well motivated to study the validity of the third
main assumption, the perfect homogeneity. Undoubtedly, at least on
small scales, large inhomogeneities exist. Although the potential
cosmological consequences of the inhomogeneities were recognized
already around the same time when the homogeneous and isotropic
models of the universe were first studied, their impact on the
global dynamics of the universe is still unknown (see e.g.\
\cite{Krasinski}). Indeed, the role of the inhomogeneities has often
been debated whenever there have been some ambiguities with
cosmological observations, for example, in the 90's when in the
determination of the age of the universe there was a discrepancy
between the implications of cosmological and astronomical
observations \cite{Futamase,Russ:1996km}.

Most recently, inhomogeneities have been invoked as the culprit for
the apparent acceleration of the expansion of the
universe\footnote{Inhomogeneities as an alternative to dark energy
were first discussed in \cite{Pascual-Sanchez:1999zr}.}, in
particular by virtue of their so-called backreaction on the metric
(for a recent discussion on the issues involved and a comprehensive
list of references, see \cite{Rasanen:2006kp}). The problems in
these approaches have been mainly of technical nature, as reducing
the symmetries of the metric rapidly complicates the calculations. A
lesser difficulty arises from the fact that different
inhomogeneities can lead to identical observations
\cite{Mustapha:1998jb}, so that not even ideal observations of light
could determine the inhomogeneities uniquely. As a consequence,
there have been two conceptually distinct ways to approach the
inhomogeneities: the first one examines the effect of the
inhomogeneities on the expansion of the universe, whereas the second
tries to determine their impact on the observations directly.

In the first approach an effective description of the universe is
constructed by averaging out the spatial degrees of freedom, i.e.\
the inhomogeneities
\cite{Rasanen:2006kp,Ellis:2005uz,Buchert:1999er,Coley:2005ei}. As a
result, one obtains averaged, effective Einstein equations which, in
addition to terms found in the usual homogeneous case, include new
terms that represent the effect of the inhomogeneities, called
backreaction in this context. It has been argued that the
backreaction might account for the accelerated expansion
\cite{Rasanen:2006kp,Buchert:1999er,Kai:2006ws}.

However, since we can only observe the redshift and energy flux of
light arriving from a given source, not the expansion rate and the
matter density of the universe nor their averages, one may wonder
how the actual observables are related to the averaged equations. To
wit, since we do not observe the expansion of the universe directly,
its acceleration is also an indirect conclusion, arising from the
fact that \textit{in the perfectly homogeneous cosmological models}
dark energy is required for a good fit. Consequently, there is no a
priori reason to assume that the acceleration would be needed in the
more general inhomogeneous models of the universe. Moreover, it is
well recognized that in general it is not correct to integrate out
constrained degrees of freedom as if they were independent. Indeed,
the fact that we can make cosmological observations only along our
past light cone makes the observable universe a constrained system.

The second approach avoids these problems by studying the effect of
the inhomogeneities directly on the observable light
\cite{Mustapha:1998jb,Celerier:1999hp}. This would be virtually
impossible in the presence of generic inhomogeneities but can be
done in some simpler models, such as the spherically symmetric
Lemaitre-Tolman-Bondi (LTB) model
\cite{Lemaitre:1933qe,Tolman:1934za,Bondi:1947av}. This model works
best in describing smooth inhomogeneities at scales of $100$
${\rm{Mpc}}$ and larger; the spherical symmetry prevents to use it
as a model for the random small scale lumpiness caused by galaxies,
as noted in \cite{Biswas:2006ub}. Indeed, the LTB model has been
used to study the effect of the smooth inhomogeneities on the
cosmological observations by several authors
\cite{Biswas:2006ub,Iguchi:2001sq,Godlowski:2004gh,Alnes:2005rw,Bolejko:2005fp,Vanderveld:2006rb,Garfinkle:2006sb,Chung:2006xh,Mustapha:1997xb,Brouzakis:2006dj};
a common conception is that these models inevitably contradict the
observed homogeneity of the large scale galaxy distribution.

Although spherical symmetry is probably an unrealistic assumption
for the entire universe, the LTB model can be regarded as describing
observations that have been averaged\footnote{This is naturally only
a crude approximation; a more correct way would be to first solve
the observables in a nonspherical model and only then average out
their angular dependence.} over the celestial sphere and it is
therefore useful at least on two counts. First, it serves as a
simple testing ground for the effect of the inhomogeneities on the
cosmological observations. Second, as the fits can be performed
unambiguously, it can be used to study the connection of the
backreaction driven effective acceleration to the observations; both
of these points will be examined in this work.

As mentioned in the beginning, two independent parameters ($H_0$ and
$\Omega_M$) are needed to uniquely define the homogeneous matter
dominated universe. In the presence of inhomogeneities, the values
of these two quantities are needed at every spatial point; that is,
the inhomogeneous dust models are defined by two functions of the
spatial coordinates: $H_0(x^i)$ and $\Omega_M(x^i)$. As a
consequence, there are inhomogeneities of two physically different
kind: inhomogeneities in the matter distribution, and
inhomogeneities in the expansion rate. Although their dynamics are
coupled via the Einstein equation, as boundary conditions they are
independent. This opens up the possibility for a universe with
inhomogeneous expansion but homogeneous present-day matter
distribution; a model of this kind could potentially fit the
supernova data as well as the galaxy surveys without invoking dark
energy.

The paper is organized as follows. Sect. \ref{ltb} discusses the
general properties of the spherically symmetric, inhomogeneous LTB
model. In Sect. \ref{models} we fit the LTB model with inhomogeneous
expansion but homogeneous present-day matter distribution to the
supernova observations. For completeness, we consider both the cases
of pure dust and dust plus the cosmological constant. The initial
conditions of these models are discussed in Sects. \ref{initialcond}
and \ref{agesect}. In Sect. \ref{averagesec} we evaluate the
expansion rate and shear in the appropriate averaged (the so-called
Buchert \cite{Buchert:1999er}) equations and demonstrate that an
accelerated average expansion is not needed to fit the supernova
data. Finally, Sect. \ref{conclusions} contains our conclusions.

\section{Spherically symmetric inhomogeneous LTB model}\label{ltb}

Since Einstein's equations form a set of six independent non-linear
second-order partial differential equations, it is impossible to
treat the universe exactly with completely generic inhomogeneities.
On the other hand, the perfectly homogeneous FRW model does not fit
the observations without a fine-tuned cosmological constant or some
other modification. Therefore, we make the next simplest
approximation and use the spherically symmetric but inhomogeneous
LTB model. The aim is to extract the leading order effects of the
inhomogeneities on the supernova observations as well as to
demonstrate the pitfalls one may encounter when using averaged
Einstein equations.

The LTB model has been used for the supernova data fitting several
times before
\cite{Iguchi:2001sq,Alnes:2005rw,Bolejko:2005fp,Vanderveld:2006rb,Garfinkle:2006sb,Biswas:2006ub,Chung:2006xh}.
However, there is a crucial physical difference between these models
and ours. Namely, the earlier works have had inhomogeneities in the
present-day matter distribution whereas we will focus on models with
inhomogeneities only in the expansion rate. Moreover, to ease the
comparison between the FRW model, familiar to all cosmologists, and
the less-known LTB model, we will rewrite the equations in a form
where the connection of the physically equivalent quantities between
these two models becomes very transparent. To introduce the new
notation, we will rederive the main results of the LTB model in
Sects. \ref{ltbmetric} and \ref{nullgeodesics}.

\subsection{The Lemaitre-Tolman-Bondi metric}\label{ltbmetric}

Let us consider a spherically symmetric dust universe with radial
inhomogeneities as seen from our location at the center. Choosing
spatial coordinates to comove ($dx^i/dt = 0$) with the matter, the
spatial origin ($x^i=0$) as the symmetry center, and the time
coordinate ($x^0 \equiv t$) to measure the proper time of the
comoving fluid, the line element takes the form
\begin{equation}\label{metric}
ds^2 = - dt^2 + \frac{(A'(r,t))^2}{1-k(r)}dr^2 + A^{2}(r,t) \left(
d\theta^2 + \sin^2 \theta d\varphi^2 \right)~,
\end{equation}
where $k(r)$ is a function associated with the curvature of
$t={\rm{const.}}$ hypersurfaces, the scale function $A(r,t)$ has
both temporal and spatial dependence, and we use the following
shorthand notations for the partial derivatives: $'\equiv
\frac{\partial}{\partial r}$ and $\dot{} \equiv
\frac{\partial}{\partial t}$. This metric was first studied by
Lemaitre \cite{Lemaitre:1933qe}, Tolman \cite{Tolman:1934za} and
Bondi \cite{Bondi:1947av}; later, it has been used in various
astronomical and cosmological contexts \cite{Krasinski}. Note that
the homogeneous and isotropic FRW metric is a special case of Eq.
(\ref{metric}), obtained in the limit: $A(r,t) \rightarrow a(t)r$
and $ k(r) \rightarrow k r^2$, where $a(t)$ is the FRW scale factor
and $k$ is the curvature constant.

The energy momentum tensor in the above defined coordinates is given
by
\begin{equation}\label{energy}
T^{\mu}_{\phantom{\mu} \nu} = - \rho_M (r,t) \delta^{\mu}_{0}
\delta^{0}_{\nu} - \rho_\Lambda \delta^{\mu}_{\phantom{\mu} \nu}~,
\end{equation}
where $\rho_M(r,t)$ is the matter density, $u^\mu =
\delta^{\mu}_{0}$ represents the components of the 4-velocity-field
of the fluid and we have kept the vacuum energy $\rho_\Lambda$ for
generality. Although the fluid is staying at fixed spatial
coordinates, it can move physically in the radial direction; this
movement is encoded in $\sqrt{g_{11}}= A'(r,t)/\sqrt{1-k(r)}$.

When Eqs. (\ref{metric}) and (\ref{energy}) are applied to the
Einstein equation, $G^{\mu}_{\phantom{\mu} \nu} = 8 \pi G
T^{\mu}_{\phantom{\mu} \nu}$, two independent differential equations
arise:
\begin{equation}\label{yht00}
\frac{\dot{A}^2+k(r)}{A^2} + \frac{2 \dot{A} \dot{A}' + k'(r)}{A A'}
= 8 \pi G (\rho_M + \rho_\Lambda)
\end{equation}
\begin{equation}\label{yht11}
\dot{A}^2 + 2 A \ddot{A} + k(r) = 8 \pi G \rho_\Lambda A^2~.
\end{equation}
The first integral of Eq. (\ref{yht11}) is
\begin{equation}\label{int11}
\frac{\dot{A}^2}{A^2} = \frac{F(r)}{A^3} + \frac{8 \pi G}{3}
\rho_\Lambda - \frac{k(r)}{A^2}~,
\end{equation}
where $F(r)$ is a non-negative function. Substituting Eq.
(\ref{int11}) into Eq. (\ref{yht00}) gives
\begin{equation}\label{yht0011}
\frac{F'}{A'A^2} = 8 \pi G \rho_M~.
\end{equation}
By combining Eqs. (\ref{yht00}) and (\ref{yht11}) we can construct
the generalized acceleration equation
\begin{equation}\label{2ndfriedmann}
\frac{2}{3} \frac{\ddot{A}}{A} + \frac{1}{3} \frac{\ddot{A}'}{A'} =
- \frac{4 \pi G}{3} (\rho_M - 2 \rho_\Lambda)~.
\end{equation}
This equation tells that the total acceleration, represented by the
left hand side, is negative everywhere unless the vacuum energy is
large enough: $\rho_\Lambda > \rho_M/2$. However, it does not
exclude the possibility of having radial acceleration
($\ddot{A}'(r,t)>0$), even in the pure dust universe, if the angular
scale factor $A(r,t)$ is decelerating enough and vice versa. Already
a simple example like this demonstrates how the very notion of the
acceleration becomes ambiguous in the presence of the
inhomogeneities \cite{Apostolopoulos:2006eg}.

The boundary condition functions $F(r)$ and $k(r)$ are specified by
the exact nature of the inhomogeneities. Their relation to the more
familiar quantities $-$ the Hubble constant $H_0$ and the density
parameter $\Omega_M$ $-$ can be recognized by comparing Eq.
(\ref{int11}) with the Einstein equation of the homogeneous FRW
model
\begin{equation}\label{frweq}
H^2(t) \equiv \frac{\dot{a}^2(t)}{a^2(t)} = H_0^2 \left[\Omega_M
\left( \frac{a_0}{a} \right)^3 + \Omega_\Lambda + (1- \Omega_\Lambda
- \Omega_M ) \left( \frac{a_0}{a} \right)^2 \right]~,
\end{equation}
where $a_0 \equiv a(t_0)$. Thus, the comparison of Eqs.
(\ref{int11}) and (\ref{frweq}) motivates us to define the local
Hubble rate
\begin{equation}\label{hupple}
H(r,t) \equiv \frac{\dot{A}(r,t)}{A(r,t)}~,
\end{equation}
and local matter density through
\begin{equation}\label{isof}
F(r) \equiv H_0^2(r) \Omega_M(r) A_0^3(r)~,
\end{equation}
with
\begin{equation}\label{lillaf}
k(r) \equiv H_0^2(r) (\Omega_M(r) + \Omega_\Lambda(r) - 1)
A_0^2(r)~,
\end{equation}
where $A_0(r) \equiv A(r,t_0)$, $H_0(r) \equiv H(r,t_0)$, and
$\Omega_\Lambda (r) \equiv {8 \pi G \rho_\Lambda}/{3 H_0^2(r)}$.
With these definitions Eq. (\ref{int11}) takes the physically more
transparent form
\begin{equation}\label{Friidman}
H^2(r,t) = H_0^2(r) \left[ \Omega_M(r) \left(\frac{A_0}{A} \right)^3
+ \Omega_\Lambda (r) + \Omega_c(r) \left(\frac{A_0}{A} \right)^2
\right]~,
\end{equation}
where $\Omega_c(r) \equiv 1- \Omega_\Lambda (r)-\Omega_M(r)$. The
difference between the conventional Friedmann equation (\ref{frweq})
and its LTB generalization, Eq. (\ref{Friidman}), is that all the
quantities in the LTB case depend on the $r$-coordinate. This is
true even for the gauge freedom of the scale function: In the FRW
case the present value of the scale factor $a(t_0)$ can be chosen to
be any positive number. Similarly, the corresponding present-day
scale function $A(r,t_0)$ of the LTB model can be chosen to be any
smooth and invertible positive function. For the rest of this paper,
we will choose the conventional gauge
\begin{equation}\label{gauge}
A(r,t_0)=r~.
\end{equation}

Although the vacuum energy density $\rho_\Lambda$ is constant, its
value in the units of critical density $\Omega_\Lambda (r) \equiv
{\rho_\Lambda}/{\rho_{{\rm{crit}}}(r)}$ is not. This is because the
critical density itself has spatial dependence:
$\rho_{{\rm{crit}}}(r) \equiv {3 H_0^2(r)}/{8 \pi G }$. The converse
is also true: if e.g.\ $\Omega_M(r) = {\rm{constant}}$, the matter
distribution $\rho_M$ itself has spatial dependence as long as
$H_0(r) \neq {\rm{constant}}$.

Integrating Eq. (\ref{Friidman}) gives
\begin{equation}\label{secondintegral}
t_0-t = \frac{1}{H_0(r)}\int_{\frac{A(r,t)}{A_0(r)}}^{1} \frac{ dx}{
x \sqrt{\Omega_M(r)  x^{-3} + \Omega_c(r) x^{-2} +
\Omega_\Lambda(r)}}~.
\end{equation}
For any space-time point with coordinates ($t,r,\theta,\varphi$),
Eq. (\ref{secondintegral}) determines the function $A(r,t)$ and all
its derivatives. Thus the metric in Eq. (\ref{metric}) is specified
and given the inhomogeneities, all the observable quantities can be
computed.

\subsection{Relation of the inhomogeneities to the observations of
light}\label{nullgeodesics}

To compare the inhomogeneous LTB model with the supernova
observations, we need an equation that relates the redshift and
energy flux of light with the exact nature of the inhomogeneities.
For this, we must study light propagation in the LTB universe. We
will again rederive the appropriate equations for notational
clarity; a more general derivation for an off-center observer can be
found in \cite{Alnes:2006pf}.

From the symmetry of the situation, it is clear that light can
travel radially, that is, there exist geodesics with $d\theta=
d\varphi=0$. Moreover, since light always travels along null
geodesics, we have $ds^2=0$. Inserting these conditions into the
equation for the line element (\ref{metric}), we obtain the
constraint equation for light rays
\begin{equation}\label{lightprop1}
\frac{dt}{du} = - \frac{dr}{du} \frac{A'(r,t)}{\sqrt{1-k(r)}}~,
\end{equation}
where $u$ is a curve parameter and the minus sign indicates that we
are studying radially \textit{incoming} light rays.

Consider two light rays with solutions to Eq. (\ref{lightprop1})
given by $t_1 = t(u)$ and $t_2 = t(u) + \lambda(u)$. Inserting these
to Eq. (\ref{lightprop1}) we obtain
\begin{equation}\label{lightprop2}
 \frac{d}{du} t_1 = \frac{d t(u)}{du} = -
 \frac{dr}{du} \frac{A'(r,t)}{\sqrt{1-k(r)}}
\end{equation}
\begin{equation}\label{lightprop3}
\frac{d}{du} t_2 = \frac{dt(u)}{du} + \frac{d \lambda (u)}{du} = -
 \frac{dr}{du} \frac{A'(r,t)}{\sqrt{1-k(r)}} + \frac{d \lambda (u)}{du}
\end{equation}
\begin{equation}\label{lightprop4}
\frac{d}{du} t_2 = - \frac{dr}{du}
\frac{A'(r,t(u)+\lambda(u))}{\sqrt{1-k(r)}} = - \frac{dr}{du}
\frac{A'(r,t) + \dot{A}'(r,t) \lambda(u) }{\sqrt{1-k(r)}}~,
\end{equation}
where Taylor expansion has been used in the last step and only terms
linear in $\lambda(u)$ have been kept. Combining the right hand
sides of Eqs. (\ref{lightprop3}) and (\ref{lightprop4}) gives the
equality
\begin{equation}\label{lightprop5}
\frac{d \lambda (u)}{du} = - \frac{dr}{du} \frac{\dot{A}'(r,t)
\lambda(u)}{\sqrt{1-k(r)}}~.
\end{equation}
Differentiating the definition of the redshift, $z \equiv
({\lambda(0)-\lambda(u)})/{\lambda(u)}$, we obtain
\begin{equation}\label{lightprop6}
\frac{dz}{du} = - \frac{d \lambda(u)}{du} \frac{\lambda
(0)}{\lambda^2(u)} = \frac{dr}{du} \frac{(1+z) \dot{A}'
(r,t)}{\sqrt{1-k(r)}}~,
\end{equation}
where in the last step we have used Eq. (\ref{lightprop5}) and the
definition of the redshift. Finally, we can combine Eqs.
(\ref{lillaf}), (\ref{lightprop1}) and (\ref{lightprop6}) to obtain
the pair of differential equations
\begin{equation}\label{dtdz}
\frac{dt}{dz} = \frac{-A'(r,t)}{(1+z) \dot{A}'(r,t)}
\end{equation}
\begin{equation}\label{drdz}
\frac{dr}{dz} =
\frac{\sqrt{1+H^2_0(r)(1-\Omega_M(r)-\Omega_\Lambda(r))A^2_0(r)}}{(1+z)
\dot{A}'(r,t)}~,
\end{equation}
determining the relations between the coordinates and the observable
redshift: $t(z)$ and $r(z)$.

Now that we have related the redshift to the inhomogeneities, we
still need the relation between the redshift and the energy flux
$F$, or the luminosity-distance, defined as $d_L \equiv \sqrt{{L}/{4
\pi F}}$, where $L$ is the total power radiated by the source. The
desired relation is given by \cite{Ellis}
\begin{equation}\label{lumdist}
d_L (z) = (1+z)^2 A(r(z),t(z))~.
\end{equation}
As the relations $t(z)$ and $r(z)$ are determined by Eqs.
(\ref{dtdz}) and (\ref{drdz}) and the scale function $A(r,t)$ by Eq.
(\ref{secondintegral}), using Eq. (\ref{lumdist}) one can calculate
$d_L$ for a given $z$. All of these relations have a manifest
dependence on the inhomogeneities (i.e.\ on the functions $H_0(r)$
and $\Omega_M(r)$). What remains is a comparison of Eq.
(\ref{lumdist}) with the observed $d_L(z)$.

In the FRW model the parameters that best describe our universe are
found by maximizing the likelihood function $e^{-\chi^2
(H_0,\Omega_M,\Omega_\Lambda)}$ constructed from the observations.
However, to find the boundary conditions of the LTB universe that
best describe our universe, we should in principle maximize the
likelihood \textit{functional} $e^{-\chi^2 [ H_0 (r) , \Omega_M (r)
]}$. In practice, this is impossible. Therefore we will consider
some physically motivated types for the functions $H_0 (r)$ and
$\Omega_M (r)$ that contain free parameters; these are then fitted
to the supernova observations by maximizing the leftover likelihood
function.

\section{Models with inhomogeneous expansion}\label{models}

In this Section, we will discuss four models where the form of the
boundary condition functions $H_0(r)$ and $\Omega_M(r)$ has been
specified and fit them to the data of the Riess et.\ al.\ gold
sample of 157 supernovae\footnote{Note that we have chosen the inner
luminosity, or the total radiation power, of type Ia supernovae as
in \cite{Riess:2004nr}.} \cite{Riess:2004nr}. An extensive
cosmological data analysis would also take into account the galaxy
surveys and the anisotropies of the cosmic microwave background.
However, we will be satisfied to give only qualitative arguments as
to why our models could have the potential to fit these data sets as
well; discussion of the potential fit for the CMB data will be given
in Sects. \ref{diskussio} and \ref{conclusions}.

According to the observations, galaxies seem to be rather evenly
distributed in the universe \cite{Eisenstein:2005su}. Indeed, the
usual objection to the inhomogeneous models is that they contradict
the observed homogeneity of the large scale structure
\cite{Vanderveld:2006rb}. However, as stressed in Sect. \ref{intro},
the matter distribution of the present universe can be homogeneous
even though the expansion rate would have spatial variations. This
motivates us to consider models with a uniform present-day matter
distribution.

When modelling the universe as perfectly homogeneous, its expansion
has to accelerate in order to fit the supernova data.
Mathematically, acceleration means that the second time derivative
of the scale function is positive, but as can be seen from Eq.
(\ref{2ndfriedmann}), this is not possible without the cosmological
constant or some other form of dark energy. But then again, the
observations are made along the past light cone, and what affects
the observations is the variation of the dynamical quantities along
the past light cone, not just the time variation. This is naturally
true in the homogeneous universe as well, but as the time variation
differs from the variation along the light cone only in the presence
of inhomogeneities, one does not usually bother to make the
distinction. However, here the difference is essential.

The directional derivative along the past light cone reads as
\begin{equation}\label{directionalderivative}
\frac{d}{dt} = \frac{\partial}{\partial t} +\frac{dr}{dt}
\frac{\partial}{\partial r} = \frac{\partial}{\partial t} -
\frac{\sqrt{1-k(r)}}{A'(r,t)} \frac{\partial}{\partial r}  \approx
\frac{\partial}{\partial t} - \frac{\partial}{\partial r}~,
\end{equation}
where the approximation in the last step is more accurate for the
small values of $r$, but is qualitatively correct even for larger
$r$. The main content of Eq. (\ref{directionalderivative}) is that
from the observational point of view, the negative $r$-derivative
roughly corresponds to the positive time derivative.  This is
natural since by looking at a source, we simultaneously look into
the past (i.e. along the \textit{negative} $t$-axis) and spatially
further (i.e. along the \textit{positive} $r$-axis). So to mimic the
acceleration, i.e.\ for the expansion rate to look like it would
increase towards us along the past light cone, the expansion
$H_0(r)$ must \textit{decrease} as $r$ grows: $H_0'(r)<0$.

With the above given arguments in mind, we have chosen the following
form for the boundary condition functions:
\begin{eqnarray}\label{boundcond}
H_0(r) &=& H + \Delta H e^{-r/r_0}~,\nonumber\\
\Omega_M(r) &=& \Omega_0 = {\rm{constant}}~,
\end{eqnarray}
where $H$, $\Delta H$, $r_0$ and $\Omega_0$ are free parameters
determined by the supernova observations and the exponential has
been chosen for simplicity. In the model of Sect. \ref{model1}, all
the four parameters are left free whereas in Sect. \ref{model2} we
have fixed $\Omega_0=1$. For generality, the cosmological constant
has been included in Sect. \ref{model3}; for computational
simplicity and to facilitate the comparison with the model of Sect.
\ref{model2} we have also set: $\Omega_M(r)+\Omega_\Lambda(r)=1.$

As explained in Sect. \ref{ltbmetric}, the present-day matter
distribution in the models with $\Omega_M(r)={\rm{const.}}$ is not
perfectly uniform since the critical density depends on $H_0(r)$.
Hence we have also studied a model with
$\rho_M(r,t_0)={\rm{const.}}$ in Sect. \ref{model4}. The discussion
of all the fits is given in Sect. \ref{diskussio} and the possible
problems with the initial conditions of these models are then
discussed in Sects. \ref{initialcond} and \ref{agesect}.

\subsection{Inhomogeneous expansion and dust with $\Omega_M(r) =
{\rm{const.}}$}\label{model1}

\begin{figure}[htp]
\begin{center}
\includegraphics[width=12.5cm]{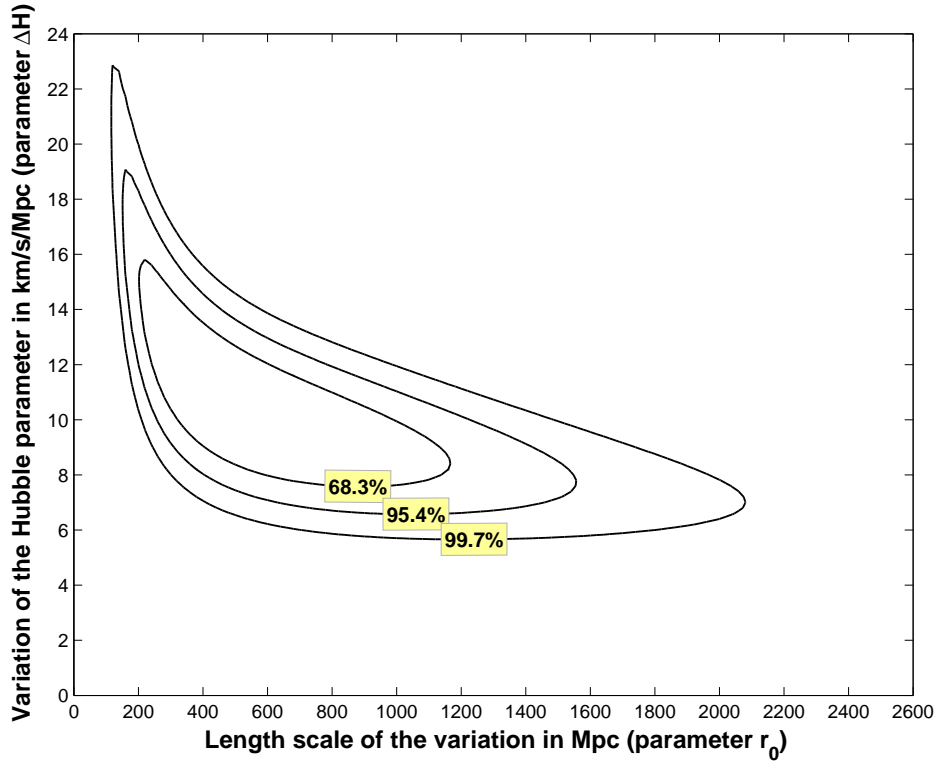}
\caption{Confidence level contours in the LTB model with
$\Omega_M(r) = {\rm{constant}} = 0.45$ and $H_0(r) =$ $56.3$
${\rm{km/s/Mpc}} + \Delta H e^{-r/r_0}$.}\label{figur1}
\end{center}
\end{figure}

For fixed values of the parameters ($H,\Delta H, r_0, \Omega_0$),
one can compute the luminosity-distance-redshift relation of Eq.
(\ref{lumdist}). By repeating this computation for different values
of the parameters we search for the maximum value of the likelihood
function $e^{-\chi^{2}(H,\Delta H, r_0, \Omega_0)}$, i.e.\ the
minimum value of
\begin{equation}\label{chisquared}
\chi^{2} \equiv \sum_{n=1}^{157} \left(
\frac{d_L^{{\rm{obs}}}(z_n)-d_L(z_n)}{\sigma_n} \right)^2~,
\end{equation}
where $d_L^{{\rm{obs}}}(z_n)$ is the observed luminosity-distance
for a source with redshift $z_n$ and $\sigma_n$ is the estimated
error of the measured $d_L^{{\rm{obs}}}$. In this way, we find that
the best fit values for the parameters in this model are:
\begin{itemize}
\item $H + \Delta H = 66.8$ ${\rm{km/s/Mpc}}$
\item $\Delta H = 10.5$ ${\rm{km/s/Mpc}}$
\item $r_0=500$ ${\rm{Mpc}}$
\item $\Omega_0 = 0.45$
\item Goodness of the fit: $\chi^2=172.6$, ${\chi^2}/{157}=1.10$
\end{itemize}
The confidence level contours with $\Omega_0$ and $H$ fixed to their
best fit values are shown in Fig. \ref{figur1}. For comparison with
the homogeneous case, the best fit nonflat $\Lambda$CDM has
$\Omega_M = 0.5$, $\Omega_\Lambda = 1.0$ and $\chi^2=175$,
$\chi^2/157=1.11$.

\subsection{Inhomogeneous expansion and dust with
$\Omega_M(r)=1$}\label{model2}

\begin{figure}[h!]
\begin{center}
\includegraphics[width=12.5cm]{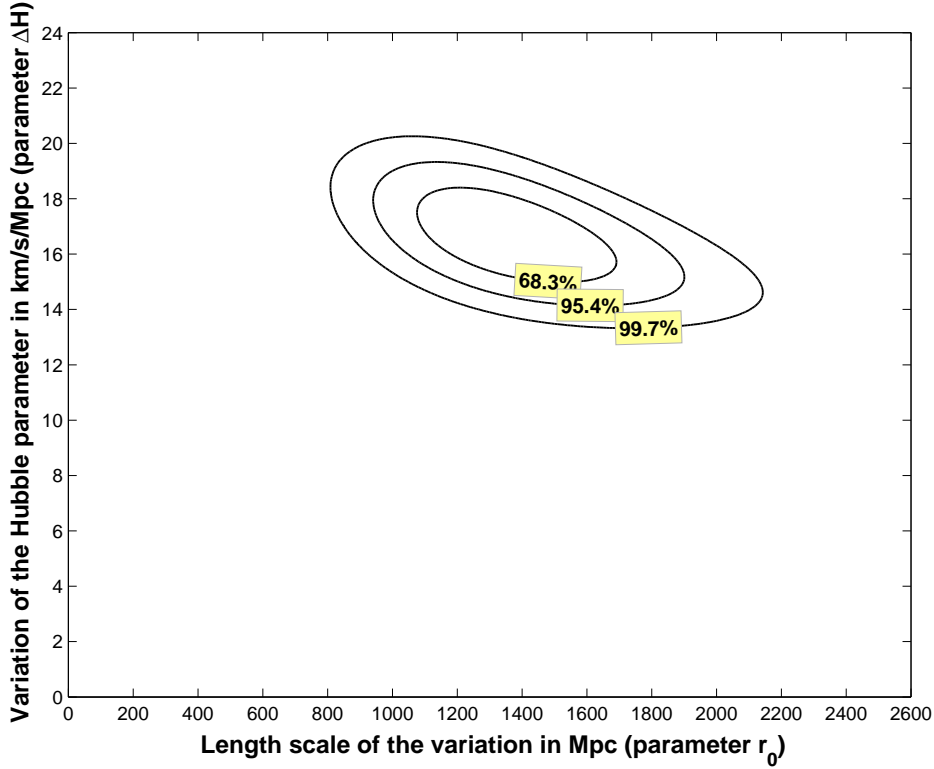}
\caption{Confidence level contours in the LTB model with
$\Omega_M(r) = {\rm{constant}} = 1$ and $H_0(r) =$ $48.7$
${\rm{km/s/Mpc}} + \Delta H e^{-r/r_0}$.}\label{figur2}
\end{center}
\end{figure}

The boundary condition functions have here the same form as in the
model of Sect. \ref{model1}, but this time we have fixed
$\Omega_M(r)=1$ in analogy with the flat FRW model. A great virtue
of this model is that the scale function can be explicitly solved
from Eq. (\ref{secondintegral}). In this case, the best fit values
for the parameters are:
\begin{itemize}
\item $H + \Delta H = 65.5$ ${\rm{km/s/Mpc}}$
\item $\Delta H = 16.8$ ${\rm{km/s/Mpc}}$
\item $r_0=1400$ ${\rm{Mpc}}$
\item Goodness of the fit: $\chi^2=176.3$, $\chi^2/157=1.12$
\end{itemize}
The confidence level contours with $H$ fixed to its best fit value
are displayed in Fig. \ref{figur2}. For comparison with the
homogeneous case, the best fit flat concordance $\Lambda$CDM model
has $\Omega_M = 0.3$, $\Omega_\Lambda = 0.7$ and $\chi^2=177$,
${\chi^2}/{157}=1.13$.

\newpage

\vspace{-0.7cm}

\subsection{Inhomogeneous expansion with
$\Omega_M(r)+\Omega_\Lambda(r)=1$}\label{model3}

We have now seen that inhomogeneous models of the universe can fit
the supernova data without the cosmological constant. On the other
hand, the homogeneous model with the cosmological constant fits the
data as well. Thus, it is natural to ask whether an even better fit
could be obtained with both the cosmological constant and the
inhomogeneities. Let us therefore require that
$\Omega_M(r)+\Omega_\Lambda(r)=1$ with the same form for the
function $H_0(r)$ as before. Then we find
\begin{itemize}
\item $H + \Delta H = 66$ ${\rm{km/s/Mpc}}$
\item $\Delta H = 8$ ${\rm{km/s/Mpc}}$
\item $r_0=600$ ${\rm{Mpc}}$
\item
$\Omega_\Lambda \equiv \rho_\Lambda/\rho_{{\rm{crit}}} =0.33$ for
fixed $\rho_{{\rm{crit}}}= 3H^2/{(8 \pi G) }$ with $H= 66$
${\rm{km/s/Mpc}}$
\item Goodness of the fit: $\chi^2=173.4$, $\chi^2/157=1.10$
\end{itemize}
Comparing this model with the homogeneous flat $\Lambda$CDM
($\chi^2/157=1.13$) and the analogous inhomogeneous dust model of
Sect. \ref{model2} ($\chi^2/157=1.12$) suggests that the mixture of
the inhomogeneous expansion and the vacuum energy does not give a
better fit than either of them separately.

\newpage

\begin{figure}[h!]
\begin{center}
\includegraphics[width=12.5cm]{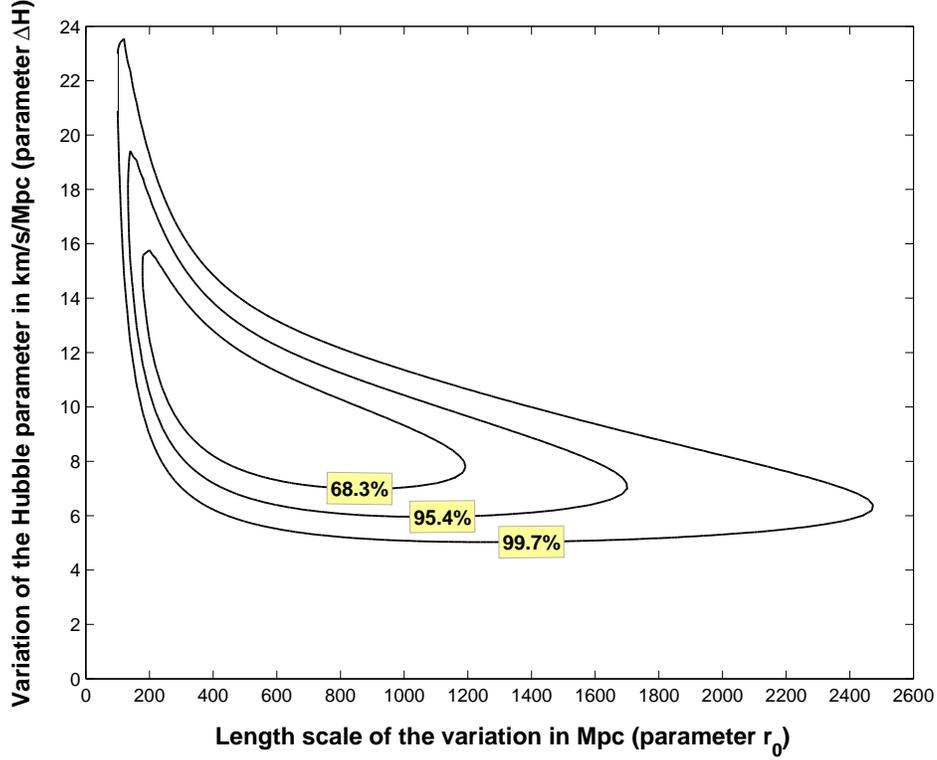}
\caption{Confidence level contours in the LTB model with perfectly
uniform present-day matter density: $H_0(r)=57$ ${\rm{km/s/Mpc}}+
\Delta H e^{-r/r_0}$, $\Omega_M(r) = 0.29 (67$ ${\rm{km/s/Mpc}}
)^2/H_0^2(r)$.} \label{figur3}
\end{center}
\end{figure}

\vspace{-0.4 cm}

\subsection{Inhomogeneous expansion and dust with
$\rho_M(r,t_0)={\rm{const.}}$}\label{model4}

Let us then consider a model with strictly uniform present-day
matter distribution: $\rho_M(r,t_0)={\rm{constant}}$. As can be seen
in Eq. (\ref{rhomatter}), the condition for the uniformity is:
$H_0^2(r) \Omega_M(r) = {\rm{constant}}$. Therefore, we choose the
boundary condition functions to be of the form
\begin{eqnarray}\label{boundcond34}
H_0(r) &=& H + \Delta H e^{-r/r_0}~,\nonumber\\
\Omega_M(r) &=& \Omega_0 (H + \Delta H)^2/(H + \Delta H
e^{-r/r_0})^2~.
\end{eqnarray}
The data analysis then gives the following best fit values:
\begin{itemize}
\item $H + \Delta H = 67$ ${\rm{km/s/Mpc}}$
\item $\Delta H = 10$ ${\rm{km/s/Mpc}}$
\item $r_0= 450$ ${\rm{Mpc}}$
\item $\Omega_0 = 0.29$
\item Goodness of the fit: $\chi^2=172.6$, ${\chi^2}/{157}=1.10$
\end{itemize}
The confidence level contours with $\Omega_0$ and $H$ fixed to their
best fit values are displayed in Fig. \ref{figur3}.

\begin{figure}
\begin{center}
\includegraphics[width=10.0cm]{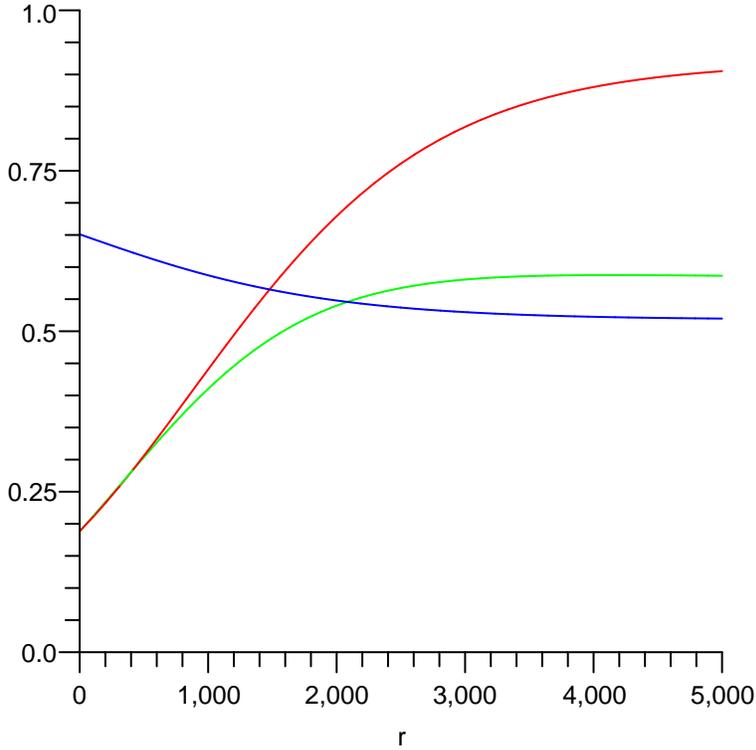}
\caption{ The functions $\Omega_M(r)$ (red line), $H_0(r)$ in the
units of $100$ ${\rm{km/s/Mpc}}$ (blue line) and the physical matter
density today $\rho_M(r,t_0)$ in the units of the critical density
at the origin $\rho_{{\rm{crit}}}(0)$ (green line) for the best fit
values in the model with simultaneous Big Bang.} \label{figur4}
\end{center}
\end{figure}

\vspace{-0.4 cm}

\subsection{Inhomogeneous expansion with simultaneous Big
Bang}\label{model5}

Finally, let us consider a model with simultaneous Big Bang, i.e. a
spatially constant age of the universe (\ref{agelt1}). This
constraint leaves us with only one free function. Hence, to maintain
the radially decreasing Hubble expansion, we choose the boundary
condition functions to have the form
\begin{eqnarray}\label{boundcond35}
H_0(r) &=& H \left[ \frac{\sqrt{1-\Omega_M(r)} - \Omega_M(r)
{\rm{arsinh}}\sqrt{\frac{1-\Omega_M(r)}{\Omega_M(r)}} }{
(1-\Omega_M(r))^{3/2}} \right] ~,\nonumber\\
\Omega_M(r) &=& \frac{\Omega_0}{(1+\delta e^{-r/r_0})^2}~.
\end{eqnarray}
The data analysis gives that the best fit values for these
parameters are: $H = 76.5$ ${\rm{km/s/Mpc}}$, $\delta = 1.21$, $r_0=
1000$ ${\rm{Mpc}}$, $\Omega_0 = 0.92$, goodness of the fit:
$\chi^2=175.5$, $\chi^2/157=1.12$. These values imply that the
Hubble function $H_0(r)$ varies from the value $H_0(0)=65$
${\rm{km/s/Mpc}}$ near us to its asymptotic value $H_0(r \gg
r_0)=52$ ${\rm{km/s/Mpc}}$, as shown in Fig \ref{figur4}. Eq.
(\ref{agelt1}) then gives the age of the universe simply as
$t_{{\rm{age}}} = 1/H = 12.8$ ${\rm{Gyr}}$. The values are similar
to the model of Ref. \cite{Alnes:2005rw}.

\subsection{Discussion of the fits}\label{diskussio}

As proved by Mustapha, Hellaby and Ellis, any isotropic set of
observations can be explained by appropriate inhomogeneities in the
LTB model \cite{Mustapha:1998jb}. Hence, the fact that we could find
boundary conditions fitting the supernova observations is not
surprising. Instead, the novel aspect here is the form of these
functions that gives the best fit, as well as their physical
interpretation: inhomogeneities in the expansion rate but
homogeneous present-day matter density. The particular interest lies
in the fact that the analysis is not only in qualitative agreement
with the observed homogeneity in galaxy surveys, but also gives a
similar value for the present-day matter density: $\Omega_M \lesssim
0.4$. Of course, the matter power spectrum should be reanalyzed in
the LTB model to confirm the quantitative agreement, but we do not
expect a small spatial variation of the Hubble parameter to change
the result notably.

Indeed, the smallness ($\sim15 \%$) of the spatial variation in the
Hubble parameter is another significant result of the fits in Sects.
\ref{model1} and \ref{model4}. The variation can be considered
small, as it is of the same order with the uncertainty of the
model-independently\footnote{Note that the smaller uncertainties
found in the CMB data analysis cannot be used here as those fits
assume that the entire universe is perturbatively close to the
homogeneous FRW model.} deduced value for the local Hubble rate
\cite{Freedman:2000cf}. The variation of the Hubble parameter found
by Alnes, Amarzguioui and Gr{\o}n \cite{Alnes:2005rw} has also
similar magnitude, but their model contains a large ($\sim 400 \%$)
variation in the matter density.

In addition to the ones discussed in Sects.
\ref{model1}-\ref{model5}, we have considered various other forms
for the boundary condition functions $\Omega_M(r)$ and $H_0(r)$. A
generic outcome is that inhomogeneities in $H_0(r)$ appear to have a
much bigger effect on the goodness of the fit than the
inhomogeneities in $\Omega_M(r)$. In fact, we have been able to
obtain good fits for the $H_0(r) = H + \Delta H e^{-r/r_0}$ -models
with inhomogeneities of almost any kind on $\Omega_M(r)$. Moreover,
we have found that models with $H_0(r)={\rm{const.}}$ do not give a
good fit irrespective of what kind of inhomogeneities we have
inserted in $\Omega_M(r)$. As an exception, a relatively high
($\Omega_0 \sim 30$) and thin ($r_0 \sim 150$ ${\rm{Mpc}}$) peak of
the form\footnote{The term $ \Omega_{0}/(3e^{4})$ has been included
to make the matter density in Eq. (\ref{rhomatter}) positive.}
$\Omega_M(r)=\Omega_0 e^{-r/r_0} + \Omega_{0}/(3e^{4})$ seems to
slightly improve the fit, giving $\chi^2 \sim 180$ when
$H_0(r)={\rm{const.}}$; the result can be understood by looking at
Eq. (\ref{rhomatter}), which places the derivatives of the functions
$\Omega_M(r)$ and $H_0(r)$ on a similar position, and as $H_0'(r)<0$
can mimic acceleration, so can $\Omega_M'(r)<0$.

The results of Sect. \ref{model3} indicate that the inhomogeneities
of the expansion rate and vacuum energy are mutually exclusive in
the sense that their combination does not lead to a better fit. A
good fit can be achieved either by having vacuum energy but no
inhomogeneities; by having inhomogeneities but no vacuum energy; or
by having both about the half amount of their separately deduced
best fit values, i.e.\ they seem to have a very similar effect on
the supernova observations.

The early supernova data had a discrepancy between the Hubble
parameter deduced from the low redshift sample and the one deduced
from the high redshift sample \cite{Padmanabhan:2002vv}, but it
seems to have vanished from the latest supernova data
\cite{Choudhury:2003tj}. However, the results of Sects.
\ref{model1}-\ref{model5} suggest that the feature still exists, and
rather than a discrepancy in the data, it is actually an alternative
explanation to dark energy. Note that the LTB model certainly
describes a universe with spatially varying expansion rate more
realistically than two separate FRW models, one for high redshift
regime and another for low redshifts. In addition, the discrepancy
in the early supernova data between the local and global Hubble
parameters had a similar magnitude with our best fit value for the
variation of the Hubble parameter. The length scale associated with
the variation of the Hubble parameter, found to lie within the range
of about $200$ to $1200$ ${\rm{Mpc}}$ in our analysis, matches the
explanation as well.

As can be seen in Sect. \ref{model2}, the explicitly solvable LTB
model with $\Omega_M(r)=1$ fits the data essentially as well as the
more complicated ones with $\Omega_M(r) \neq 1$. Hence it will be a
useful example when discussing the time evolution of the
inhomogeneities in Sect. \ref{initialcond} and the average expansion
in Sect. \ref{averagesec}.

In order to respect the cosmological principle, we should not live
in the dead center of a region that is expanding faster than the
global average. It was shown in \cite{Alnes:2006uk} that the
supernova data does not impose severe restrictions for the location
of an off-center observer, but as argued in \cite{Alnes:2006pf},
perhaps the most relevant constraint comes from the dipole
anisotropy of the CMB. Here we give a rough estimate for the
off-center distance allowed by the CMB dipole.

We define the effective peculiar velocity, caused by the
inhomogeneous expansion, that an observer at the coordinate $r=d$
has relative to the symmetry center
\begin{equation}\label{offcenter1}
v_p(d) = d (H_0(0)-H_0(d)),
\end{equation}
which simply measures the deviation from the Hubble law. In a
homogeneous universe, the effective peculiar velocity of Eq.
(\ref{offcenter1}) is identically zero and the observed peculiar
velocity is accounted for motion relative to the comoving
coordinates. We instead assume this coordinate velocity negligible
and require that the effective peculiar velocity of Eq.
(\ref{offcenter1}) gives rise to the CMB dipole. Inserting the
velocity deduced from the CMB dipole, $v_p = 370$ ${\rm{km/s}}$
\cite{Fixsen:1996nj}, and the boundary condition function $H_0(r)$
of Eq. (\ref{boundcond34}) with the best fit values $r_0 = 450$
${\rm{Mpc}}$, $\Delta H = 10$ ${\rm{km/s/Mpc}}$ to Eq.
(\ref{offcenter1}) gives $d=140$ ${\rm{Mpc}}$, which is somewhat
more than was found in a more thorough analysis for a different
model\footnote{In \cite{Alnes:2006pf} the function $H_0(r)$ was
fixed up to the distance of the LSS, whereas we have left its tail
to be fixed from the CMB observations. Thus, instead of $\Delta H$,
we use $H_0(0)-H_0(d)$ in Eq. (\ref{offcenter1}).}
\cite{Alnes:2006pf}. This is a non-negligible fraction of the size
of the inhomogeneity, $d/r_0 = 0.3$, so according to this naive
estimate the CMB dipole does not constrain our position strictly to
the center. Of course even greater off-center distances are allowed
if we would also have coordinate velocity in the opposite direction
to counterbalance the effective peculiar velocity of Eq.
(\ref{offcenter1}).

\subsection{Time evolution of the inhomogeneities}\label{initialcond}

In the preceding discussion we defined the models by giving the
boundary conditions $H_0(r)$ and $\Omega_M(r)$ on the present-day
spatial hypersurface. However, as is inherent in the Big Bang
cosmology, the question of the "naturalness"\ of the initial
conditions arises in models with inhomogeneous expansion but a
homogeneous present-day matter distribution. In this section, we
address the problem by considering the time evolution of the
inhomogeneities in the explicitly solvable LTB model with
$\Omega_M(r)=1$. We expect the same qualitative features to hold in
the more general case with $\Omega_M(r) \lesssim 1$ as well. For
more on the subject, see e.g. \cite{Krasinski}.

Combining Eqs. (\ref{yht0011}) and (\ref{isof}) we obtain the
following expression for the matter density:
\begin{equation}\label{rhomattertime}
\rho_M (r,t) = \rho_M (r,t_0) \left(
\frac{A_0^2(r)A_0'(r)}{A^2(r,t)A'(r,t)} \right)~,
\end{equation}
where
\begin{equation}\label{rhomatter}
\rho_M (r,t_0) = \frac{3H_0^2(r)}{8 \pi G} \Omega_M(r) \left[1 +
\frac{A_0(r)}{3A_0'(r)} \left( \frac{\Omega_M'(r)}{\Omega_M(r)}
 + 2 \frac{H_0'(r)}{H_0(r)} \right) \right]~.
\end{equation}
With $\Omega_M(r)=1$ and $A_0(r)=r$, Eq. (\ref{rhomattertime})
reduces to
\begin{equation}\label{rhomattertime2}
\rho_M(r,t) = \frac{3H_0^2(r) + 2r H_0'(r)H_0(r)}{8\pi
G[1+\frac{3H_0(r)}{2}(t-t_0)][r(t-t_0)H_0'(r)+ (1+
\frac{3H_0(r)}{2}(t-t_0))]}~,
\end{equation}
which gives the explicit time dependence of the matter distribution.
Using Eq. (\ref{rhomattertime2}), one finds that the ratio of the
matter density near us ($r=0$) to its asymptotic value ($r = R \gg
r_0$) is
\begin{equation}\label{rhoratio}
\frac{\rho_M(0,t)}{\rho_M(R,t)} = \left[ \frac{H_0(0)}{H_0(R)}
\frac{\left(1+\frac{3}{2}H_0(R)(t-t_0)\right)}{\left(1+\frac{3}{2}H_0(0)(t-t_0)\right)}
\right]^2~.
\end{equation}
At late times, $t = T \gg t_0$, Eq. (\ref{rhoratio}) gives
\begin{equation}\label{rhofuture}
\frac{\rho_M(0,T)}{\rho_M(R,T)} \longrightarrow 1~,
\end{equation}
which means that, irrespective of the initial conditions, the
homogeneity of the matter distribution is an attractor solution.

By performing the integral of Eq. (\ref{secondintegral}), one can
calculate the explicit time dependence of the expansion rate:
\begin{equation}\label{Timedependenthubble}
H(r,t) = \frac{H_0(r)}{1 + \frac{3H_0(r)}{2}(t-t_0)}~.
\end{equation}
Eq. (\ref{Timedependenthubble}) gives the ratio of the local ($r=0$)
and global ($r = R \gg r_0$) Hubble parameters
\begin{equation}\label{hubbleratio}
\frac{H(0,t)}{H(R,t)} = \frac{H_0(0)}{H_0(R)}
\frac{\left(1+\frac{3}{2}H_0(R)(t-t_0)\right)}{\left(1+\frac{3}{2}H_0(0)(t-t_0)\right)}~,
\end{equation}
which tells us that the inhomogeneity of the expansion rate will
also vanish in the late ($t = T \gg t_0$) universe:
\begin{equation}\label{hubblefuture}
 \frac{H(0,T)}{H(R,T)} \longrightarrow 1~.
\end{equation}

By comparing Eqs. (\ref{rhoratio}) and (\ref{hubbleratio}) one sees
that the inhomogeneity of the matter density will diminish more
rapidly than the inhomogeneity of the expansion rate. This explains
in a natural way why, with generic initial conditions, the present
universe would have negligible inhomogeneities in the matter
distribution but detectable inhomogeneities in the expansion rate,
like our best fit model in Sect. \ref{model4}. On the other hand, as
noted in Sect. \ref{diskussio}, even if today's variations in the
matter density were of the same order with the required small
variation in the expansion rate, the observations might not be able
to detect them.

Turning the argument around, Eqs. (\ref{rhoratio}) and
(\ref{hubbleratio}) tell us that in order to have noticeable
inhomogeneities today, the early universe should have been very
inhomogeneous. Naturally, we do not have an explanation for this as
there is no theory that would determine the initial conditions of
the universe. However, when considering the backward time evolution,
one should keep in mind that the dust approximation will break down
at some stage and as a consequence, Eqs. (\ref{rhomattertime2}) and
(\ref{Timedependenthubble}) do not represent the inhomogeneities of
the very early universe.

\subsection{The age of the universe}\label{agesect}

The age of the perfectly homogeneous universe is a well-defined
number, the difference $t_{{\rm{age}}} \equiv t_0-t_{BB}$ of the
time coordinate today $t_0$ and the time coordinate for which the
scale function goes to zero $ a(t \rightarrow t_{BB}) \rightarrow
0$. However, with inhomogeneities the notion of the age of the
universe becomes more complicated as the age can depend on the
spatial location (see e.g. \cite{Krasinski}). In the LTB model this
can be seen by noting that the singularity condition $A(r,t_{BB})=0$
now defines a curve $t_{BB}(r)$ on the $(r,t)$-plane whereas in the
homogeneous FRW case $a(t_{BB})=0$ defines a unique point $t_{BB}$
on the $t$-axis. We give the explicit expressions of the age of the
LTB universe as a function of the boundary conditions $H_0(r)$ and
$\Omega_M(r)$ for different cases in Appendix \ref{age}.

The expressions (\ref{agelt1}), (\ref{agegt1}), (\ref{ageeq1}) and
(\ref{ccageeq1}) are similar to the homogeneous FRW case, except for
the $r$-dependence. In this case, the \textit{local} value of the
Hubble parameter gives the order of magnitude for the \textit{local}
age of the universe. Thus, if $\Omega_M(r) = {\rm{const.}}$, already
small spatial fluctuations in the Hubble parameter $H_0(r)$ lead to
quite big variations in the age of the universe between different
spatial locations. For a constant $\Omega_M(r)$ the scale of the
spatial variation in the age of the universe is simply given by
$(\Delta H)^{-1}$, and at scales of ${\cal O}(1000)$ ${\rm{Mpc}}$
would thus be some billions of years for the best fit models of
Sects. \ref{model1}-\ref{model4}. Although theoretically
inhomogeneous Big Bang could be a viable possibility (one need only
to think of e.g. colliding branes; see
\cite{Kallosh:2001ai,Erickson:2006wc}), age differences of such
great magnitude could be hard to reconcile with any existing model
of the very early universe. They certainly contradict the spirit of
inflation, where the whole observable universe arises from a single
causal domain. However, the possibility remains that, for some
reason, $H_0(r)$ and $\Omega_M(r)$ are related in such a fashion
that they give $t_{{\rm{age}}}(r) = {\rm{constant}}$. The
inhomogeneous LTB model would continue to fit the supernova data
well as demonstrated by the model of Sect. \ref{model5}.

\section{Is accelerated average expansion needed?}\label{averagesec}

The expansion of the dust dominated LTB universe does not accelerate
anywhere but nevertheless can, as shown e.g.\ in Sect. \ref{models},
fit the supernova observations. However, it has been argued that a
certain kind of average expansion can accelerate even though the
local expansion would decelerate everywhere
\cite{Rasanen:2006kp,Buchert:1999er,Kai:2006ws}. The aim in this
section is to study if this kind of average acceleration appears in
our best fit models of Sect. \ref{models}. For this purpose, we need
the Buchert equations that describe the averaged dynamics of a
general irrotational dust universe \cite{Buchert:1999er}
\begin{equation}\label{buchert}
3 \frac{\ddot{a}(t)}{a(t)} = -4 \pi G \langle \rho_M \rangle_D +
\frac{2}{3} ( \langle \theta^2 \rangle_D - \langle \theta
\rangle_D^2 ) - \langle \sigma^{\mu \nu} \sigma_{\mu \nu} \rangle_D
\end{equation}
\begin{equation}\label{buchertt}
3 \frac{\dot{a}^2(t)}{a^2(t)} = 8 \pi G \langle \rho_M \rangle_D -
\frac{1}{2} \langle R^{(3)} \rangle_D - \frac{1}{3} ( \langle
\theta^2 \rangle_D - \langle \theta \rangle_D^2 ) + \frac{1}{2}
\langle \sigma^{\mu \nu} \sigma_{\mu \nu} \rangle_D
\end{equation}
\begin{equation}\label{bucherttt}
\frac{\partial}{\partial t} \langle \rho_M \rangle_D + 3
\frac{\dot{a}(t)}{a(t)} \langle \rho_M \rangle_D = 0~,
\end{equation}
where $\sigma^{\mu \nu} \sigma_{\mu \nu} \geq 0$ represents the
shear, $a(t) \equiv (\int_D \sqrt{ {\rm{det}} [g_{ij}]} d^3
x)^{1/3}$ is the averaged scale factor, $D$ is the integration
domain, $g_{ij}$ is the spatial part of the metric, $R^{(3)}$ is the
curvature scalar of the $t={\rm{const.}}$ spatial hypersurface,
$\theta \equiv \nabla_\mu u^\mu$ is the expansion scalar and the
spatial average of a scalar $S$ is defined as
\begin{equation}\label{average}
\langle S \rangle_D \equiv \frac{\int_D S \sqrt{ {\rm{det}}
[g_{ij}]} d^3 x}{\int_D \sqrt{ {\rm{det}} [g_{ij}]} d^3 x }~.
\end{equation}
The difference of Buchert's acceleration equation (\ref{buchert})
and its homogeneous FRW counterpart is known as the backreaction
\begin{equation}\label{bakreaktion}
Q \equiv \frac{2}{3} ( \langle \theta^2 \rangle_D - \langle \theta
\rangle_D^2 ) - \langle \sigma^{\mu \nu} \sigma_{\mu \nu} \rangle_D
~.
\end{equation}

The average expansion accelerates if the right hand side of Eq.
(\ref{buchert}) is positive; this can be achieved by having large
enough variance of the expansion rate although it is counterbalanced
by the average shear. The variance gets large values when
contracting ($\theta<0$) and expanding ($\theta>0$) regions coexist,
and in fact the average acceleration has been connected to
gravitational collapse \cite{Rasanen:2006kp,Apostolopoulos:2006eg}.
However, as demonstrated in \cite{Paranjape:2006cd}, a globally
expanding dust universe can have average acceleration as well.

The definition of the shear tensor is
\begin{equation}\label{shear}
\sigma_{\mu \nu} \equiv \frac{1}{2} ( \nabla_\mu u_\nu + \nabla_\nu
u_\mu ) - \frac{1}{3} ( g_{\mu \nu} + u_\mu u_\nu) \nabla_\alpha
u^\alpha ~ ,
\end{equation}
where $u^\mu$ is the four-velocity of the dust. Calculating the
shear and the expansion scalar for the LTB metric in the coordinates
of Eq. (\ref{metric}) gives
\begin{equation}\label{shear2}
\sigma^{\mu \nu} \sigma_{\mu \nu} = \frac{2}{3} \left(
\frac{\dot{A}}{A} - \frac{\dot{A}'}{A'} \right)^2
\end{equation}
\begin{equation}\label{expscalar}
\theta = 2 \frac{\dot{A}}{A} + \frac{\dot{A}'}{A'} \equiv 2H(r,t) +
H_r(r,t)~.
\end{equation}
Using the field equation (\ref{int11}) one finds that the two Hubble
functions $H(r,t)$ and $H_r(r,t)$, defined in Eq. (\ref{expscalar}),
are related at the present time\footnote{In general, $t=t_0$ is the
moment when the boundary conditions $\Omega_M(r)$ and $H_0(r)$ have
been specified.}  $t_0$ as
\begin{equation}\label{hubbler}
H_r(r,t_0) = H_0(r) + r H_0'(r)~,
\end{equation}
so that at $t=t_0$ the shear and the expansion rate have the
expressions
\begin{eqnarray}\label{shearnow}
\sigma^{\mu \nu} \sigma_{\mu \nu}(t=t_0) &=& \frac{2}{3} (rH_0'(r))^{2}\nonumber\\
\theta(t=t_0) &=& 3 H_0(r) + rH_0'(r)~.
\end{eqnarray}
Inserting Eqs. (\ref{boundcond34}) and (\ref{shearnow}) to Buchert's
acceleration equation (\ref{buchert}), we obtain the exact
expression for the average acceleration of the model in Sect.
\ref{model4}:
\begin{equation}\label{buchert2}
3 \frac{\ddot{a}(t_0)}{a(t_0)} = - \frac{3}{2} \Omega_0 H^2 + 6
\langle H_0^2 \rangle + 4 \langle r H_0' H_0 \rangle - 6 \langle H_0
\rangle^2 - 4 \langle r H_0' \rangle \langle H_0 \rangle -
\frac{2}{3} \langle r H_0' \rangle^2~.
\end{equation}
We choose the integration domain in Eq. (\ref{buchert2}) as the
origin-centered ball with radius $R$, taken to be greater than the
inhomogeneity scale (e.g.\ $R \approx 2 r_0$) but smaller than the
horizon distance $R < 1/H$. Note that at $t=t_0$ we have $\sqrt{
{\rm{det}} [g_{ij}]} = r^2 / \sqrt{1-k(r)}$ in the integration
measure. As the curvature function of Eq. (\ref{lillaf}) is small
for the models of Sect. \ref{models}, we can make the linear
approximation, $1/ \sqrt{1-k(r)} \approx 1 + k(r)/2 $, for the
integrals in Eq. (\ref{average}) to get an approximate expression
for the average of a scalar $S$
\begin{equation}\label{average2}
\langle S \rangle_{\mathbb{B}(R)} \equiv \frac{\int_0^{R} S r^2
(1-k(r))^{-1/2 }dr}{\int_0^{R} r^2 (1-k(r))^{-1/2} dr } \approx
\langle S \rangle_c + \frac{1}{2} \langle S k \rangle_c -
\frac{1}{2} \langle S \rangle_c \langle k \rangle_c ~,
\end{equation}
where we have used $\langle S \rangle_c$ to denote the "coordinate"\
average:
\begin{equation}\label{average3}
\langle S \rangle_c \equiv \frac{\int_0^{R} S r^2 dr}{\int_0^{R} r^2
dr }~.
\end{equation}
Using the approximation of Eq. (\ref{average2}), we can calculate
the integrals of Eq. (\ref{buchert2}) to obtain an expression for
the average acceleration of the model in Sect. \ref{model4}:
\begin{equation}
\begin{split}\label{average4}
\frac{3}{r_0^2 H^4} \frac{\ddot{a}(t_0)}{a(t_0)} = - \frac{3
\Omega_0}{2 r_0^2 H^2} + \bigg( \frac{64}{81} h^2 + \frac{9}{2} h -
288 (1-\Omega_0) \bigg) \frac{h^2}{x^3}
e^{-x} \\
+ \bigg(  \frac{6}{5} (1-\Omega_0) x^2 + 9
(1-\Omega_0) x + 135 (1-\Omega_0) \frac{1}{x} + 279 (1-\Omega_0) \frac{1}{x^2} \\
- \frac{27}{4} \frac{h^2}{x^3} - 432 \frac{h}{x^3} +\frac{567}{2} (1-\Omega_0) \frac{1}{x^3} + 42 (1-\Omega_0) \bigg) h^2 e^{-2x} \\
+ \bigg( 2 x^2 +\frac{43}{3} x + \frac{1855}{9} \frac{1}{x} + \frac{11405}{27} \frac{1}{x^2} +\frac{69223}{162} \frac{1}{x^3} + \frac{586}{9} \bigg) h^3 e^{-3x}  \\
+ \bigg( \frac{1}{2} x^2 + \frac{53}{24} x + \frac{2945}{288}
\frac{1}{x} + \frac{19475}{1728} \frac{1}{x^2} +
\frac{124313}{20736} \frac{1}{x^3} +\frac{419}{72} \bigg) h^4
e^{-4x}
 \\
- \frac{9}{256} \frac{h^4}{x^3} +\frac{16}{81} \frac{h^3}{x^3}
+\frac{9}{2} (1-\Omega_0) \frac{h^2}{x^3}~,
\end{split}
\end{equation}
where we have denoted $h \equiv \Delta H/ H$ and $x \equiv R/r_0$. A
careful inspection reveals that the negative first term on the right
hand side of Eq. (\ref{average4}) dominates, at least for the values
of the parameters $r_0$, $\Delta H$, $H$ and $\Omega_0$ that give a
good fit. Thus, according to Eq. (\ref{average4}), the average
expansion does not accelerate in the best fit model of Sect.
\ref{model4}. The validity of Eq. (\ref{average4}) is verified by
the fact that it actually deviates only about $10^{-7}$ from the
numerically computed accurate result of the integrals in Eq.
(\ref{buchert2}) for the best fit values $h=10/57$, $r_0=450$
${\rm{Mpc}}$ and $\Omega_0=0.4$. We have repeated the computation of
the average expansion also for the model of Sect. \ref{model1} and
found no acceleration there either.

Finally, let us consider the LTB model with $\Omega_M(r)=1$. For
this model, both the backreaction of Eq. (\ref{bakreaktion}) and the
3-curvature $R^{(3)}$ vanish identically \cite{Paranjape:2006cd}.
Thus, the Buchert equations (\ref{buchert}), (\ref{buchertt}) and
(\ref{bucherttt}) reduce to the FRW equations of the flat and
homogeneous dust universe. However, a creature living in the LTB
universe would observe the luminosity-redshift relation calculated
from the exact Einstein equations, not from the FRW equations. The
difference between these results is significant since, as shown in
Sect. \ref{model2}, the LTB model with $\Omega_M(r)=1$ fits the
supernova observations whereas the flat matter dominated FRW model
does not \cite{Riess:2004nr}. As the averaged matter density
$\langle \rho_M \rangle_D$ has only a small dependence on the
averaging domain $D$ in our models, the conclusion would not change
even if one were to use a different smoothing scale $R_i$ for every
supernova at $r=R_i$. Overall, there seems to be no meaningful way
of using the averaged equations to calculate the actual observables,
e.g. the luminosity-redshift relation.

\section{Conclusions}\label{conclusions}

We have studied the effect of inhomogeneities on the cosmological
supernova observations in the spherically symmetric matter dominated
LTB universe. Our main conclusions are as follows:

\begin{enumerate}
\item The inhomogeneous matter dominated LTB model fits the supernova data better than the homogeneous
FRW model with the cosmological constant.

\item Inhomogeneities of the expansion rate seem to have much bigger effect on the
goodness of the fit than the inhomogeneities of the matter
distribution.

\item The model giving the best fit has perfectly uniform present-day matter
density with $\Omega_M \sim 0.3$, in agreement with the galaxy
surveys, but its Hubble parameter varies from the local value of
$\sim 70$ ${\rm{km/s/Mpc}}$ to the value of $\sim 60$
${\rm{km/s/Mpc}}$ over the distance scale of $\sim500$ ${\rm{Mpc}}$
from us.

\item Neither local nor average acceleration is needed for a good fit to the supernova
data.

\item At least in the LTB universe, the averaged Buchert equations give
a notably different prediction for the observable
luminosity-redshift relation than the exact Einstein equations.
\end{enumerate}

The first of the above listed results is neither new nor surprising,
as it was proved in \cite{Mustapha:1998jb} that any isotropic set of
observations can be explained by appropriate inhomogeneities, and
later demonstrated with examples by several authors. Instead, the
second result is a new one, as is the model that explains both the
observations of supernovae and the observed homogeneity in the large
scale distribution of galaxies without invoking dark energy.
Moreover, we found in Sect. \ref{initialcond} that within the
dynamics of the Einstein equation, the universe may evolve to this
kind of a configuration without any substantial amount of
fine-tuning in the initial conditions. In addition, as the
supernovae and galaxy surveys constrain the boundary condition
functions $H_0(r)$ and $\Omega_M (r)$ only for the small values of
the radial coordinate $r$, the tails of these functions can still be
freely chosen to give a good fit for the CMB power spectrum as well.
In fact, Alnes, Amarzguioui and Gr{\o}n have realized this in
practice by fitting the LTB model to the first peak of the CMB
\cite{Alnes:2005rw}. It has also been shown that even a homogeneous
model can fit the CMB data without dark energy
\cite{Blanchard:2003du}.

Our another focus was on the averaged dynamics of the dust universe,
given by the Buchert equations that illustratively demonstrate the
potential cosmological significance of the inhomogeneities. In Sect.
\ref{averagesec}, we showed that none of our dust models that fit
the supernova observations have accelerated average expansion. On
the other hand, one can easily construct examples that have average
acceleration but are ruled out by observations, whereas dust models
that would both have the average acceleration and fit the
cosmological data are still to be presented. Moreover, as discussed
at the end of Sect. \ref{averagesec}, it seems that the Buchert
equations cannot be used to calculate the observable
luminosity-redshift relation. Although explicitly demonstrated only
in the LTB universe, we expect that it might be the case in the more
general models as well; after all, the Buchert equations are
obtained by integrating over a spatial hypersurface whereas the
observables are calculated by integrating over the past light cone.
Altogether, averaging over a spatial hypersurface is perhaps not
practical if one wants to make comparison with the observations.
Instead, probably a better alternative is to relate the
inhomogeneities directly to the observations, as done e.g.\ in this
paper and, if needed, take the averages only of the relevant
observables, such as the energy flux and redshift of light.

Of course, some sort of smoothing is implicitly assumed also in the
LTB model. Therefore, the study of observations in the LTB universe
does not answer the interesting question about the potential
cosmological consequences of the small scale lumpiness caused by
galaxies. Their effect could be significant, possibly depending also
on how smoothly dark matter, perhaps the dominant energy component
of the universe, is distributed in the universe.


\acknowledgments{We thank Maria Ronkainen for checking the
calculations as well as for useful comments on the manuscript. We
also thank Syksy R\"as\"anen, Tomi Koivisto, Filippo Vernizzi, Subir
Sarkar, Thomas Buchert, Martin Kunz, Morad Amarzguioui and
H{\aa}vard Alnes  for helpful comments and discussions. T.M.\ is
supported by Helsinki University Science Foundation.}

\appendix

\newpage

\section{The age of the LTB universe}\label{age}

The age of the LTB-universe can be calculated by integrating the
field equation (\ref{Friidman}). In the cases where the result is an
elementary function of the boundary conditions, the results are:

\begin{enumerate}

\item $\Omega_M(r) < 1$ and $\Omega_\Lambda (r) = 0$:
\begin{equation}\label{agelt1}
t_{{\rm{age}}}(r) = \frac{\sqrt{1-\Omega_M(r)} - \Omega_M(r)
{\rm{arsinh}}\sqrt{\frac{1-\Omega_M(r)}{\Omega_M(r)}} }{H_0(r)
(1-\Omega_M(r))^{3/2}}~.
\end{equation}

\item $\Omega_M(r) > 1$ and $\Omega_\Lambda (r) = 0$:
\begin{equation}\label{agegt1}
t_{{\rm{age}}}(r) = \frac{\Omega_M(r)
{\rm{arcsin}}\sqrt{\frac{\Omega_M(r)-1}{\Omega_M(r)}} -
\sqrt{\Omega_M(r)-1} }{H_0(r) (\Omega_M(r)-1)^{3/2}}~.
\end{equation}

\item $\Omega_M(r) = 1$ and $\Omega_\Lambda (r) = 0$:
\begin{equation}\label{ageeq1}
t_{{\rm{age}}}(r) = \frac{2}{3 H_0(r)}~.
\end{equation}

\item $\Omega_M(r) + \Omega_\Lambda (r) = 1$:
\begin{equation}\label{ccageeq1}
t_{{\rm{age}}}(r) = \frac{2}{3 H_0(r)}
\frac{{\rm{arsinh}}\sqrt{\frac{1-\Omega_M(r)}{\Omega_M(r)}}}{\sqrt{1-\Omega_M(r)}}~.
\end{equation}

\end{enumerate}

\end{document}